# A Reflective Gaussian Coronagraph for ExAO: Laboratory Performance

Ryeojin Park

College of Optical Sciences, University of Arizona, Tucson, AZ 85721

Laird M. Close, Nick Siegler, Eric L. Nielsen and Thomas Stalcup

Steward Observatory, University of Arizona, Tucson, AZ 85721

rpark@email.arizona.edu

## ABSTRACT

We report laboratory results of a coronagraphic test bench to assess the intensity reduction differences between a "Gaussian" tapered focal plane coronagraphic mask and a classical hard-edged "Top Hat" function mask at Extreme Adaptive Optics (ExAO) Strehl ratios of ~94%. However, unlike a traditional coronagraph design, we insert a reflective focal plane mask at 45° to the optical axis. We also used an intermediate secondary mask ("Mask_2") before a final image in order to block additional mask-edge diffracted light. The test bench simulates the optical train of ground-based telescopes (in particular the 8.1m Gemini North telescope). It includes one spider vane, different mask radii (r= 1.9$\lambda$/D, 3.7$\lambda$/D, 7.4$\lambda$/D) and two types of reflective focal plane masks (hard-edged "Top Hat" and "Gaussian" tapered profiles). In order to investigate the relative performance of these competing coronagraphic designs with regard to extra-solar planet detection sensitivity, we utilize the simulation of realistic extra-solar planet populations (Nielson et al. 2006). With an appropriate translation of our laboratory results to expected telescope performance, a "Gaussian" tapered mask radius of 3.7$\lambda$/D with an additional mask ("Mask_2") performs best (highest planet detection sensitivity). For a full survey with this optimal design, the simulation predicts ~30% more planets detected compared to a similar sized "Top Hat" function mask with "Mask_2." Using the best design, the point contrast ratio between the stellar PSF peak and the coronagraphic PSF at 10$\lambda$/D (0.4″ in H band if D = 8.1m) is ~10 times higher than a classical Lyot "Top Hat" coronagraph. Hence, we find a Gaussian apodized mask with an additional blocking mask is a superior (~10x higher contrast) than use of a classical Lyot coronagraph for ExAO–like Strehls.

*Subject heading:* instrumentation: adaptive optics --- (stars:) planetary systems

## 1. Introduction

Numerical simulations of different coronagraphic designs have been studied extensively in the last few years (for example, Malbet 1996; Itoh et al. 1998; Sivaramakrishnan et al. 2001; Kuchner & Traub 2002). These simulations tried to find optimized coronagraphic designs which suppress bright starlight while minimizing the inner working angle (IWA). In order to detect the direct light from extra-solar giant planets (EGPs), these designs are required to suppress starlight more than a million times at tenths of arcsecond of separation (Burrows et al. 2004). These efforts to detect direct light from extra-solar planets become more feasible as recent adaptive optics (AO) systems are producing diffraction-limited images at near-infrared wavelengths with large 8m-class ground based telescopes. Furthermore, the next generation of AO systems such as Gemini's GPI (Macintosh et al. 2006) and an ESO's Very Large Telescope (VLT) SPHERE (Dohlen, K. et al., 2006) will operate in the regime of "extreme adaptive optics" (ExAO). These ExAO systems are designed to achieve Strehl ratios of ~ 95% in H band. Although a few coronagraphs (for example, the CIAO AO camera at the Subaru telescope (Iye et al. 2004), and the NACO system at the VLT (Lenzen at al. 2003)) are deployed on ground based AO telescopes, their delivered Strehl ratios are typically still much less than ~80%. NICI camera at the Gemini South (Toomey & Ftaclas 2003) is planned to start in 2007 following instrument commissioning. The Lyot Project coronagraph should test ExAO Strehl ratios regularly at the telescope once the AEOS high order AO system is fully optimized (Sivaramakrishnan et al.

2004).

In this study, we present laboratory results comparing how different coronagraph masks work at these ExAO Strehl ratios (~94%), and how much contrast can be achieved with real systems in the lab. By performing test bench experiments, the current real-world fabrication constraints of this set of masks and stops can also be tested before deployment. We hope our contribution compliments the theoretical work of Lloyd and Sivaramakrishnan (2005) and Sivaramakrishnan et al. (2001) by empirical test of different coronagraph at ExAO Strehls.

In this paper, we compare a classical (Lyot 1939) hard-edged "Top Hat" masks to "Gaussian" tapered focal plane masks. We choose a reflective focal plane mask design since we believe it is optimal in feeding a future high speed tip-tilt sensor (or an interferometric focal plane WFS of Codona & Angel 2004). We investigate the optimized designs and compare relative performance using three different size of masks (r=1.9$\lambda$/D, 3.7 $\lambda$/D, 7.4 $\lambda$/D). We also test and quantify the additional intensity reduction of inserting an intermediate secondary occulting mask ("Mask_2") into the beam between a final image and a science camera. In order to evaluate the relative performance of different coronagraphic designs, we utilize the extra-solar planet population simulation of Nielsen et al. 2006.

## 2.   Optical Setup

### 2.1.   The fabrication of the mask

Both "Gaussian" and "Top Hat" profile focal plane masks (Fig. 1) were fabricated by Precision Optical Imaging in Rochester, New York, USA. The pseudo-Gaussian masks were built to the specifications shown in Figure 2. The specifications were re-designed in the desired half tone value patterns and converted into AutoCAD format. Using this format, the high reflection (HR) chrome pixels (small squares) are closely controlled in size. All dot sizes are 10 μm and of thickness 0.4 μm. The spacing between dots is varied corresponding to the designed opacity. Using microlithographic techniques, HR chrome is deposited onto a common B270 optical glass (2 mm thick, no AR coating allowed). The B270 glass is used because of good transmission properties in the visible light. The flatness of the glass substrate is better than 1 λ/square inch. The "science" beam transmits through the coronagraph mask plate. To obtain the actual transmission profiles, we scanned the focal plane masks with a photo scanner at 2400 dpi resolution and present them in Figure 3. The sum of transmission, absorption and reflection was normalized to one. The combination of reflection and absorption losses from the glass substrate of the mask was about 4% (typical for non-AR coated glass). The elliptical mask shape (shown in Figs. 1& 2) allows the masks to be inserted 45° to the optical axis where they act as a "mini fold flat" reflecting cores of the PSF for a future high speed tilt sensor (or an interferometric focal plane WFS of Codona & Angel 2004). With the mask positioned at 45°, the incident beam views a circular mask in projection (see Fig.1 (b)).The main purpose of the WFS2 is very high speed tilt control of beam to align the beam on the mask (see for example, Angel et al. 2006), so the spatial filtering of the small mask does not affect measured tilt. Neither does it affect its use as a reference beam to feed for an interferometric focal plane WFS.

## 2.2. Optical setup

The optical system has been designed using the optics design program ZMAX and is presented in Figure 4(a). The test bench is located in an optical laboratory at Steward Observatory, University of Arizona. The layout of our optical train setup and its photograph is shown in Figure 4(b). The optical train is designed to simulate the Gemini North 8.1m telescope. It does so by scaling Gemini's 8.1m entrance pupil to $7.00 \pm 0.05$ mm and its 1.18m central obscuration to $1.02 \pm 0.05$ mm. The test bench used a f/90 beam with a 0.532 μm laser at the first and second focal planes, which maps to the same spot size as would be used in H band (1.6 μm) in an actual design at f/32 on a 8.1m telescope.

We used a monochromatic 0.532 μm laser beam to simulate a very narrow bandwidth filter. We will include a broad band white light source in a future study to quantify the loss of contrast with bandwidth. A compact combination of a positive achromat and a negative singlet is used in a Barlow configuration to obtain a compact but slow f/90 focal ratio (see Fig. 4(a)). Although our coronagraph system consists of λ/4 optics, we only illuminated the small central portion of each optic, so we expect ∼ λ/20 performance. Optics are laser-grade optimized lenses with /078 HEBBAR AR coating.

While we also wanted to include the diffracting effects of the secondary's two spider vanes, the actual 1 cm width vanes are too narrow for direct scaling (1.2 μm) to our 7mm pupil. We decided to proceed with only one vane with the minimum thickness produced from photochemical etching (∼100 μm) as shown in Figure 5. It allows the PSF with and without spider vanes to be simultaneously measured. However, we only analyzed the PSF profile in directions away from the diffraction spikes to ease analysis and minimize any other effects (polarization?) of our over-sized spider vanes.

The incident plane wave from the stellar source is simulated by the laser beam (doubled

Nd:YAG) passing through a 10 μm pinhole spatial filter and then collimated. To avoid saturation, neutral density filters are inserted between the laser and the objective lens as needed. The collimated beam passes through the Gemini telescope pupil simulator and focuses onto the 45° focal plane mask (focal plane #1). The beam is centered on a focal plane mask within 1 μm accuracy based on minimizing transmission through the Lyot stop.

The focal plane masks are mounted on x,y,z stages with a set of different mask sizes deposited on the same glass substrate. This allows the selection of a different mask by moving only in the x, y directions. Barlow #2 recollimates the beam and forms a pupil on the Lyot stop. The diameter of the Lyot stop is $5.49 \pm 0.05$mm (undersized by 22% in diameter) and the diameter of its central obscuration is $1.40 \pm 0.05$mm (oversized by 37% in diameter). Thus, ~41% of the pupil area is blocked after passing through the Lyot stop. All the results in this paper were obtained with this Lyot stop in the beam. For comparison, both telescope simulator and the Lyot stop are presented in Figure 5. Finally, Barlow #3 refocuses the f/90 beam onto the camera (focal plane #2). The CCD camera used is an Apogee Alta U4000 (2048x2048 CCD) with $7.4x7.4 \ \mu m^2$ pixels. The thermoelectric cooler temperature was set at -18±1 °C; CCD data flow was controlled by MaxIm DL software.

"Mask_2" is inserted between the final imaging Barlow #3 and the science camera. This extra mask (not part of a "classic" Lyot coronagraph) is designed to block the diffracted "spot of Arago", or "Poisson's spot", formed by light diffracting around the focal plane mask. We used an extra spider vane mask (0.010±0.005mm) as "Mask_2" by inserting it in the same orientation as the Lyot stop spider vane and the telescope simulator.

### 3.  Laboratory results

Testing of the laboratory coronagraph commenced on August 10, 2005. The full width at half maximum (FWHM) of the PSF in focal plane #1 was 48.1μm. The Strehl ratio at the first focal plane was reduced to 94% to simulate an "ExAO" Strehl ratio. The Strehl ratio was reduced by micro-roughness of our lenses which gave high order aberrations and array of "super-speckles" (see Fig 6(b)). Slight misalignment of optics gave low order aberration (see Fig. 6(b) and 6(c)). The Strehl ratio was determined by measuring peak count of an experimental PSF divided by its total flux, divided by the peak count of a "perfect" (theoretical) PSF divided by its total flux. To create the "perfect" PSF, we performed the forward fast Fourier transform of the autocorrelation of an over-sampled uniformly illuminated telescope pupil image from our telescope simulator pupil mask. Figure 6(a) shows the theoretical PSF image (left) and the acquired "ExAO" PSF image at the lab in focal plane #1 (right). The residual wavefront errors are a mix of low spatial frequencies and high spatial frequencies roughly similar to that expected from non-common path residual AO aberrations. The residual wavefront errors have been estimated by subtracting our theoretical PSF image from the 94% Strehl ratio lab image (see Fig. 6 (b)). We roughly obtained ExAO like "super speckle" patterns in the PSF as shown in Fig. 6(b) ;however, we caution that this is not a true ExAO PSF and so the real ExAO PSF and coronagraphic performance will be somewhat different from what we find here. We have estimated that the integrated rms wavefront error larger than spatial scales of 216 cm (assuming D= 8m), corresponding to less than the 3.7 λ/D mask size, is 60% of the total 26nm of wavefront error. The power larger than 108 cm is 90% and larger than 50cm (typical DM control scale) is 95%. The rms wavefront error figure is shown in Fig. 6(c). While we realize our PSF is not identical to that of a perfect ExAO system; however, it is a reasonable approximation of a high Strehl AO PSF.

Four different test bench setups were empirically tested: 1) we inserted both a "Gaussian" profile focal plane mask with "Mask_2"; 2) only the "Gaussian" profile focal plane mask.; 3)a "Top Hat" focal plane mask with "Mask_2"; 4) only a "Top Hat" focal plane mask. Each test bench setup was tested using different mask sizes with an interesting range, r = 1.9 λ/D, 3.7 λ/D, and 7.4 λ/D. Because of a slight instability in the laser source intensity (stable to 1% over 1 hour while data were taken), we took a mask-out PSF and a mask-in (coronagraphic) PSF consecutively. The mask-out PSF radial profile from each test bench setup is slightly different mainly because "Mask_2" is still inserted for the aforementioned 1) and 3) setups. Another possible contribution to variation is the slight micro-roughness variation of the glass substrate around each different mask. All images were taken with the same Lyot stop in the beam.

Due to the camera's limiting dynamical range (~16 bit), three images with different exposure times were taken and scaled appropriately to make one deep image with high dynamical range (~$10^7$). In the same position, "sky" images (laser off) were taken with the same exposure time. These "sky" images contained the bias noise and any stray background light. All images were properly "sky" subtracted and flat fielded. When saturated, the Apogee CCD "bleeds" in the direction of the "spider vane" diffraction spike; however, all saturated pixels were replaced with scaled up non-saturated data. The final images were profiled at a randomly selected position angle (PA=23° well away from the spider as shown in Fig 8(b)), averaged in a five pixel box around the cut profile, and plotted as radial profiles.

We did not use an azimuthally averaged profile because of the diffraction light from the spider vane. The radial profiles of each of the four test bench setups are presented in Figure 7[E1].

In Figure 7, the solid line represents the coronagraphic PSF (mask-in) profile and the dotted line represents PSF (mask-out) profile. The vertical lines in the radial profile represent the mask radius. It was not easy to directly compare the edge of a "Gaussian" mask with that of the "Top Hat" mask because the edge of a "Gaussian" mask is not clearly defined as shown in Figures 1, 2, and 3. However, for purpose of comparison, we define that a "Gaussian" mask radius is the same as the half width at half maximum (HWHM) of the radial profile.

Figure 8 shows images from the test bench using the "Gaussian" tapered mask radius of $3.7\lambda/D$. In Figure 8(a) the left image is the coronagraphic images without "Mask_2" and in Figure 8(b) the left image shows the images with "Mask_2." Note the decrease in the diffracted light at the edge of the mask.

To measure a mean "contrast ratio", we smoothed the line of the logarithmic radial profile data. The data at two standard deviations was clipped to create an average profile. The profile data (both for the non-coronagraphic PSF and the coronagraphic image) were considered only past the mask's radius + $0.8\lambda/D$. For fitting, we used the onedspec parameter (NOAO IRAF data reduction package) to get the best fit line. Lastly, the PSF (mask-out) profile was divided by the corresponding coronagraphic (mask-in) profile to obtain the final mean "contrast ratio". The "point contrast ratio" was defined as the peak value of the PSF to the corresponding coronagraphic image value at $5\lambda/D$ and $10\lambda/D$. Table 1 summarizes our results

---

[E1] NOTE TO EDITOR: Each figure 7(a), (b), (c) should appear in a separate page in print.

for the four different test bench setups. Figure 9 shows the mean contrast as a function of a mask size and the point contrast at 10$\lambda$/D as a function of a mask size. The values in Table 1 are plotted in Fig 9.

## 4. Discussion

We have completed the design and optical alignment of a coronagraphic test bench and have obtained preliminary results summarized in Table 1. The laboratory experiment carried out have allowed us to examine the intensity reduction differences between a reflective "Gaussian" tapered focal plane mask and a reflective hard-edged "Top Hat" focal plane mask.

### 4.1 Observed gains with "Mask_2"

We have also probed the effect of adding "Mask_2." This technique blocks the additional edge-diffracted light ("spot of Arago" or "Poisson's spot") after the Lyot stop's central obscuration. Interestingly, we found that by inserting this additional mask, there was further intensity reduction as shown in Figures 7 and 9. The effect of inserting "Mask_2" seems to be most significant with the largest focal plane mask (r=7.4$\lambda$/D) case. Note the bright diffracted halo around the mask in Figure 8 which occurred when "Mask_2" was omitted. However, the coronagraphic profile is partially attenuated when using "Mask_2."

### 4.2 Compared to theory

Numerical simulations of the Gemini 8.1m telescope show that intensity suppressions by a factor of ~3 to ~10 are predicted theoretically for r = 4$\lambda$/D focal plane #1 mask and a 25% undersized Lyot stop (Sivaramakrishnan et al 2001). These results are in good agreement with our result (the mean contrast we observed was ~10) with a "Gaussian" tapered mask radius of 3.7$\lambda$/D and a Lyot stop undersized by 22% in diameter. Furthermore, we observed further reduction by a factor of ~3 using "Mask_2" with the same focal plane mask. To confirm this result, we will perform detailed numerical simulations in a future paper.

### 4.3   Ranking systems with simulated extra-solar planets

In order to rank the competing coronagraphic designs with regard to predicted success in actual detecting extra-solar planets, we use the simulations of extra-solar planet populations from Nielsen et al. 2006. We begin by translating our laboratory results (Figure 7) to expected telescope performance. The angle $\lambda/D$ is converted to a 0.04″ angular separation for an 8.1m aperture at H band (1.6 µm) and we convert the coronagraphic PSF to the achievable contrast of the system. We assume that we can only detect planets down to an H magnitude of 23 ($5\sigma$ in 2 hours) and the complete ExAO system will include a form of simultaneous differential imaging (SDI) device to remove speckle noise after the coronagraph. SDI is a technique to exploit the strong methane feature in young, massive extra-solar planets so as to achieve larger contrasts between planets and their host stars (see Lenzen et al. 2004; Biller et al. 2004; Close et al. 2005). With the typical on sky (Biller et al. 2006) 10x suppression of the speckles using SDI, we estimate that with 2 hours of exposure time we can achieve H=23 at $5\sigma$ level.

Based on data from the VLT examined in Biller et al. 2004, SDI can achieve contrasts a factor of ~10 above the residual PSF, for young, methane-rich objects. Since we seek at least $5\sigma$ detections (a factor of five in the opposite direction), the end result is to divide the curves of Figure 7 (assuming these contrasts are equally applicable to H-band) by a factor of 2, and use these as the maximum $5\sigma$ contrasts between a host star and a detectable planet for 2 hours exposure time. We plot the four estimated contrast curves that we consider in our simulations in Figure 10.

The planet population simulation procedure is described at length in Nielsen et al. 2006; however, we summarize the technique briefly here. A list of the existing youngest, closest 106 stars to the Sun is assembled for a direct imaging survey for young, extra-solar planets. For each target star, we simulate 100,000 planets, randomly assigning each one orbital elements and mass, based on extrapolations of planets detected by radial velocity surveys. Assuming ages of the target stars, the mass of each planet is converted into an H-band magnitude using the giant planet models of Burrows et al. (2003). With the orbital elements and the distances to the target stars, we solve for the angular separation between planets and stars. At this point, the contrast curve (maximum contrast ratio between host and planet star detectable as a function of angular separation) is used to determine what fraction of these simulated planets can be detected. For the four contrast curves given in Figure 10, we show the results of the simulation for a typical target star in Figure 11. For each target star, the fraction of planets detected (blue points in Fig. 11) represents the probability of detecting a planet for that particular target star. If we repeat our simulations for each of the 106 target stars and then sum the detected fractions for each target, the result is the expected number of planets detected at the end of the survey. This provides a simple metric to evaluate the relative performance of different coronagraphic designs with respect to their intended purpose in an ExAO system- detecting extra-solar planets. Figure 12 shows the number of planets detections as a function of the size of the direct imaging survey (it is assumed that the survey is designed such that the best target stars are observed first). From this analysis, it is clear that the "Gaussian" tapered mask radius of $3.7\lambda/D$ with "Mask_2" performs best for a full survey, detecting an additional two planets above the closest competitor (9 vs. ~7). The simulation predicts that an additional three planets would be detectable above the "Top Hat" focal plane mask with "Mask_2" design (9 vs. ~6). Its point contrast ratios at $5\lambda/D$ and $10\lambda/D$ are ~10 times higher than the other design. In summary, the larger higher contrast r= $7.4\lambda/D$ mask blocks too many planets and a smaller $1.9\lambda/D$ mask allows too much diffracted light to pass. Therefore, in this simulation a $3.7\lambda/D$ mask is optimal for finding planets.

Table 1.   The summary of the four different test bench setup results

| | "Gaussian" mask with an axicon | | | "Top Hat" mask with an axicon | | |
|---|---|---|---|---|---|---|
| Mask radius($\lambda$/D) | 7.4 | 3.7[a] | 1.9 | 7.4 | 3.7 | 1.9 |
| Mean contrast ratio | 85 ± 2 | 28 ± 2 | 6 ± 2 | 34 ± 2 | 9 ± 3 | 11 ± 3 |
| Point contrast ratio at 5$\lambda$/D | - | 1.2x10$^5$ | 3.5x10$^4$ | - | 1.9x10$^4$ | 6.2x10$^4$ |
| Point contrast ratio at 10$\lambda$/D | 6.1x10$^6$ | 1.4x10$^6$ | 1.3x10$^5$ | 1.0x10$^6$ | 1.1x10$^5$ | 1.6x10$^5$ |
| | "Gaussian" mask without an axicon | | | "Top Hat" mask without an axicon | | |
| Mask radius ($\lambda$ /D) | 7.4 | 3.7 | 1.9 | 7.4 | 3.7 | 1.9 |
| Mean contrast ratio | 31 ± 3 | 12 ± 4 | 5 ± 3 | 46 ± 4 | 9 ± 4 | 9 ± 2 |
| Point contrast ratio at 5$\lambda$/D | - | 2.1x10$^4$ | 1.6x10$^4$ | - | 7.9x10$^3$ | 9.1x10$^3$ |
| Point contrast ratio at 10$\lambda$/D | 1.6x10$^5$ | 8.0x10$^4$ | 3.0x10$^4$ | 7.4x10$^4$ | 1.0x10$^5$ | 3.2x10$^4$ |

[a]This combination of a 3.7$\lambda$/D "Gaussian" mask with "Mask_2" was found to be the optimal choice for exoplanet detection (see section 4)

## Figure Legends

**Fig. 1.**— (a) Photo-scans of the "Gaussian" mask (left) and the "Top Hat" mask (right). The largest "Gaussian" mask has a characteristic radius of 7.4 $\lambda$ /D (356μm). The boundary of "Gaussian" mask is fuzzy because of the thinner deposit of HR Chrome. The largest "Top Hat" mask has a radius of 7.4λ/D (356μm). In contrast, the boundary of "Top Hat" mask is hard-edged. (b) A diagram of a focal plane mask deposited on the BK270 glass substrate. When inserted at 45°, the mask appears circular in projection.

**Fig. 2.**— The "Gaussian" tapered mask schematic. The profile has an elliptical shape because the mask will be inserted at 45° to the optical axis. The characteristic diameter Y is the same as the hard-edged "Top Hat" mask diameter. The schematic is not to scale.

**Fig. 3.**— The transmission radial profile of the "Gaussian" tapered mask (the solid line) and the hard-edged "Top Hat" mask (the dashed line) in FWHM. The corresponding mask radii are 1.9λ/D, 3.7λ/D, 7.4λ/D in (a), (b) and (c) where D is its full value (not the undersized "D" of the Lyot stop).

**Fig. 4.**— (a) The test bench in the lab and its layout of the reflective coronagraph for ExAO. The Gemini telescope simulator has D= 7.00 mm in diameter with a 1.02 mm central obscuration. This D scales to the 8.1 m Gemini telescope. (b) The optical system layout using the optical design program ZMAX.

**Fig.5.**— Lyot stop (left) and telescope simulator made from 0.004 inch stainless steel and photochemical etched to tolerances of <30μm. The diameter (D) of the Lyot stop (22% undersized) is 5.49 ± 0.05mm and the diameter of its central obscuration is 1.40 ± 0.05mm

(37% oversized). The diameter of telescope simulator is 7.00 ± 0.05mm. Its central obscuration is to 1.02 ± 0.05mm.

**Fig. 6.**— (a) The theoretical (100% Strehl ratio) PSF model of focal plane #1(left) and the acquired PSF CCD image (94% Strehl ratio) at the lab in focal plane #1(right). A Strehl ratio of 94% was picked to roughly simulate an ExAOC Strehl ratio. Stretch is logarithmic. (b) The rms wavefront errors obtained by the subtracting the theoretical PSF image (100% Strehl ratio) from the acquired PSF image (94% Strehl ratio). The achieved pattern roughly approximates the scattered light from an ExAO system with some super speckles. Log stretch. (c) The log-log plot of the rms wavefront error in nm as a function of aberration size scale/D. D represents a diameter of an aperture.

**Fig. 7.**— The logarithmic relative PSF intensities as a function of distance from the PSF's peak. These profiles are sampled at arbitrary chosen PA=23° and the intensity values plotted are an average of 5 pixels around the cut profile (see Fig. 8 for cut direction). The solid line represents the coronagraphic PSF (mask-in) radial profile and the dotted line represents the non-coronagraphic PSF (mask-out) radial profile. All images use the 22% undersized diameter Lyot stop. The vertical solid lines in the radial profile represent the mask radii. (a) is the result using a mask radius of 1.9 λ/D, (b) uses a mask radius of 3.7 λ/D and (c) uses the largest mask radius of 7.4 λ/D.

**Fig. 8.**— Comparison of coronagraphic CCD images and their non-coronagraphic images. (a) The left image is the coronagraphic image with the "Gaussian" tapered mask radius 3.7 λ /D and no "Mask_2" and the right image is the PSF image with no "Mask_2." (b) The left image is the coronagraphic image with the "Gaussian" tapered mask radius 3.7 λ/D with "Mask_2."

The image on the right is the PSF with the same size mask. The red-line represents the cut of position angle (PA=23°) sampled for the radial profiles. Linear stretch.

**Fig. 9.**— The contrast ratio as a function of mask size. *Top:* the mean contrast as a function of mask size. *Bottom:* the point contrast at $10\lambda/D$ as a function of mask size.

**Fig. 10.** — The contrast curves used in the simulation of extra-solar planet populations, for four coronagraph designs from Figure 7. The minimum flux of a detectable planet ($5\sigma$) is given as a function of radius, assuming an 8m telescope (where $\lambda/D=0.04''$ at 1.6μm). Contrasts are given both as magnitudes (left side) and as linear attenuation (right side).

**Fig 11.** — Simulation results for a typical target star, for each of the four coronagraph designs used. Each point represents one of 100,000 simulated planets, with mass plotted against instantaneous projected separation from the parent star. The solid curve is the contrast curve of Figure 10, the dashed line is the minimum flux limit detectable for the given exposure time, while the dashed-dot line shows the maximum planet mass that still retains a strong enough methane signature for the SDI technique to be effective. The blue points bounded between these three constraints are the planets that can be detected by the system, with the red points showing undetected planets.

**Fig. 12.** — Expected results from a survey using each of the four coronagraph designs considered in the simulation. For 106 real target stars, 100,000 planets are simulated, with the fraction that can be detected with the given system corresponding to the probability of successfully detecting a planet around that target star. The targets are ordered by that detection probability and the probability is summed for surveys of various sizes. In this comparison, the "Gaussian" tapered mask (r=3.7 $\lambda$/D) with "Mask_2" performs the best for any size of survey, detecting two additional planets beyond any other system, if all target stars are surveyed.

Fig.1

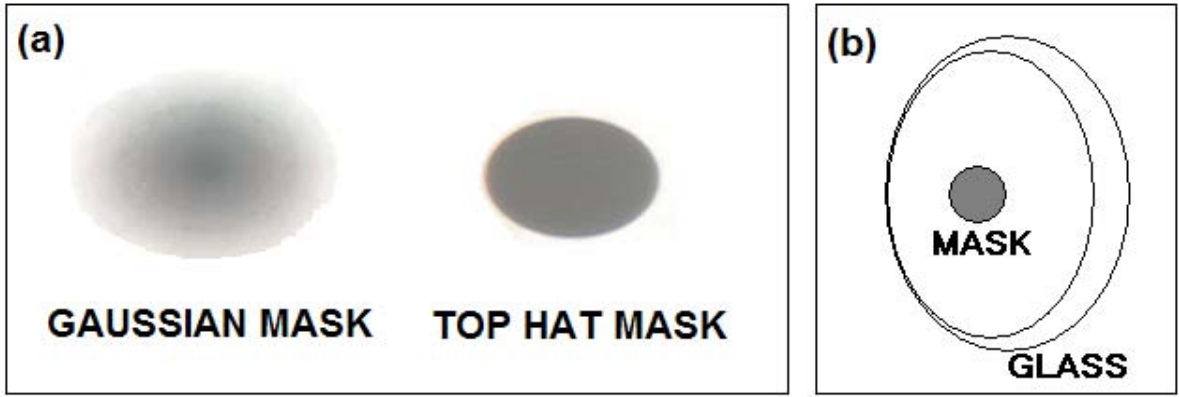

Fig.2

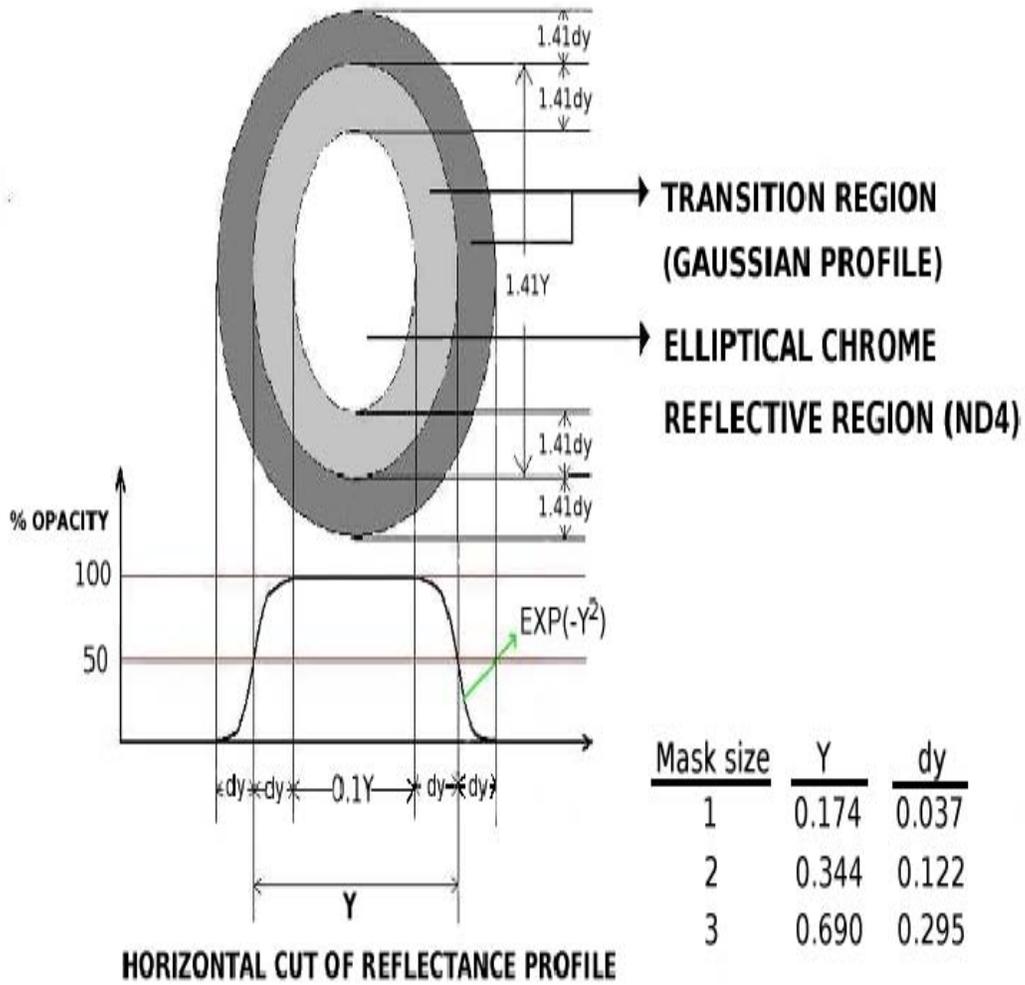

**GAUSSIAN TAPERED MASK**
**(All measurement in mm)**

| Mask size | Y | dy |
|-----------|-------|-------|
| 1 | 0.174 | 0.037 |
| 2 | 0.344 | 0.122 |
| 3 | 0.690 | 0.295 |

Fig. 3(a)

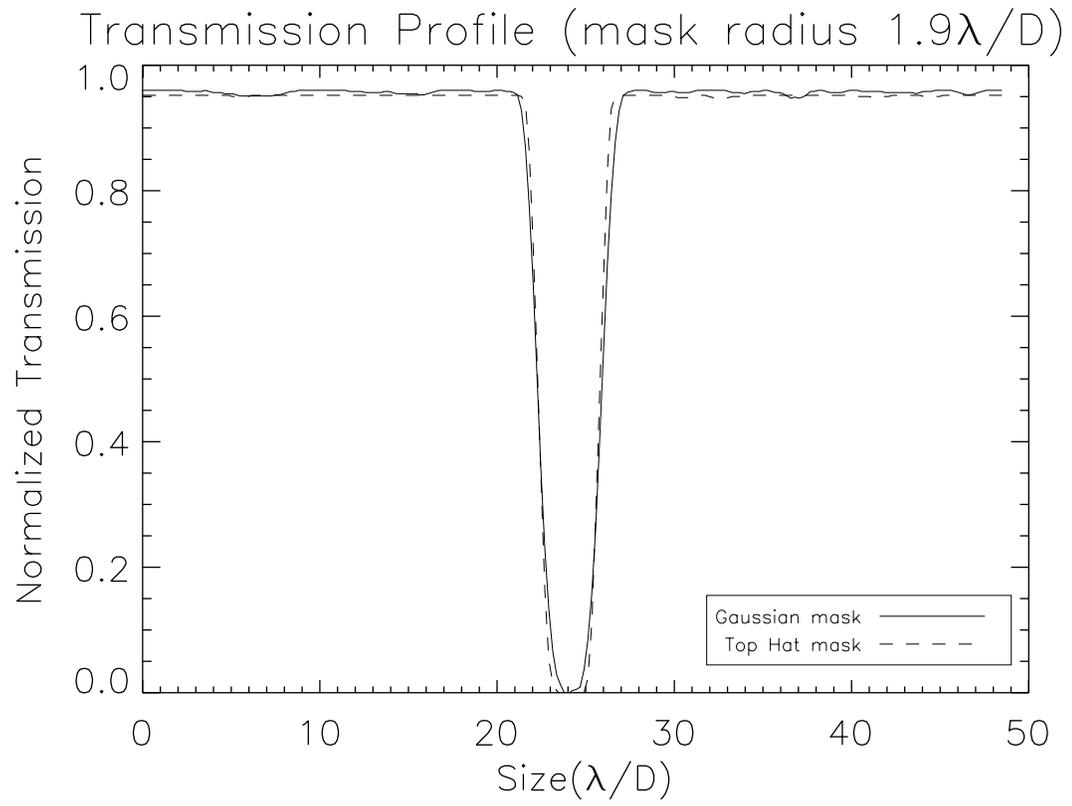

Fig. 3(b)

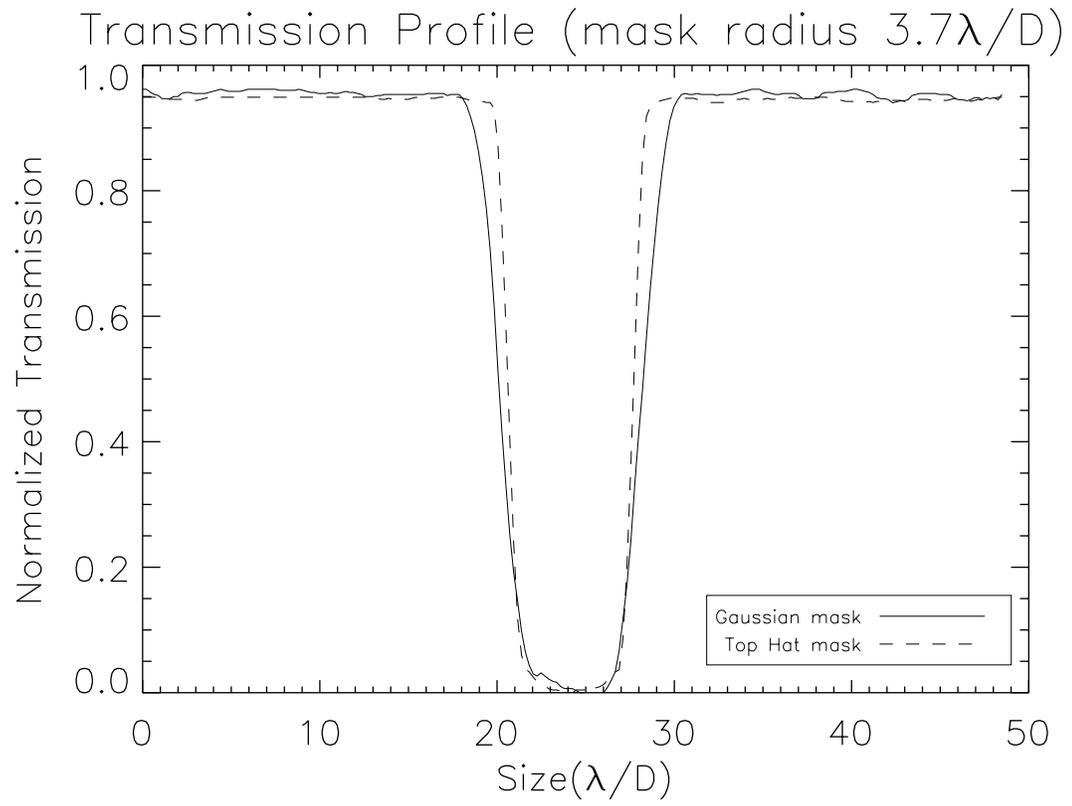

Fig. 3(c)

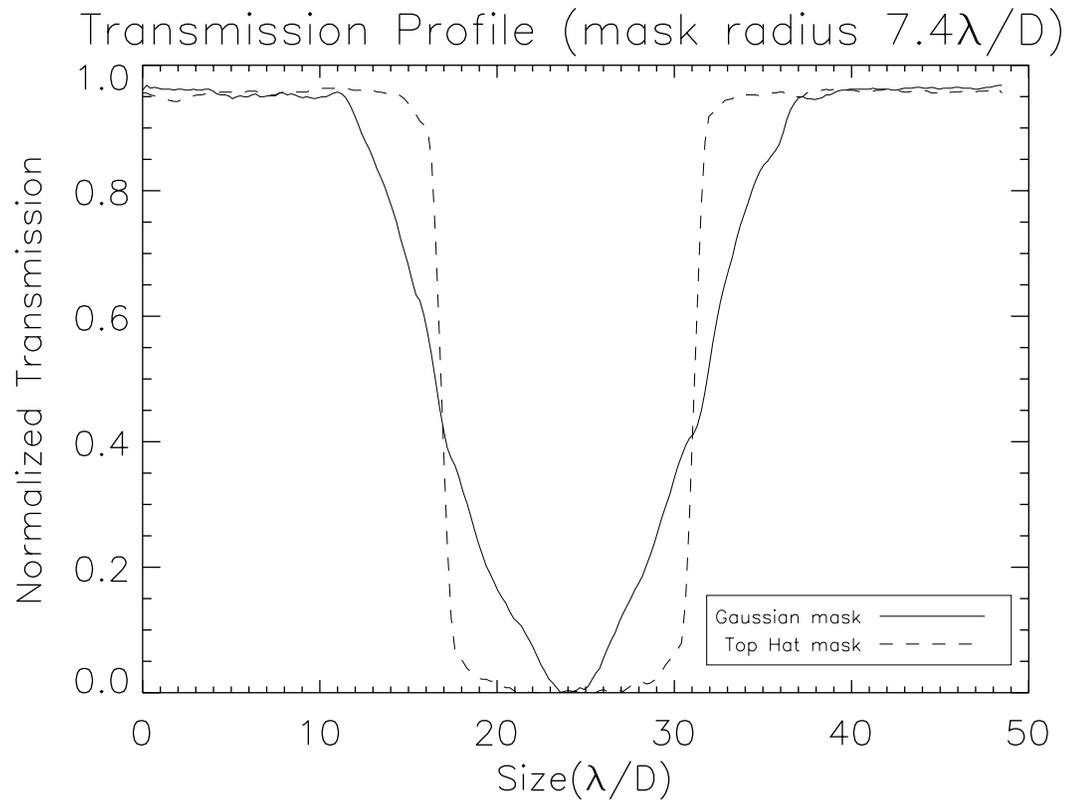

Fig.4(a)

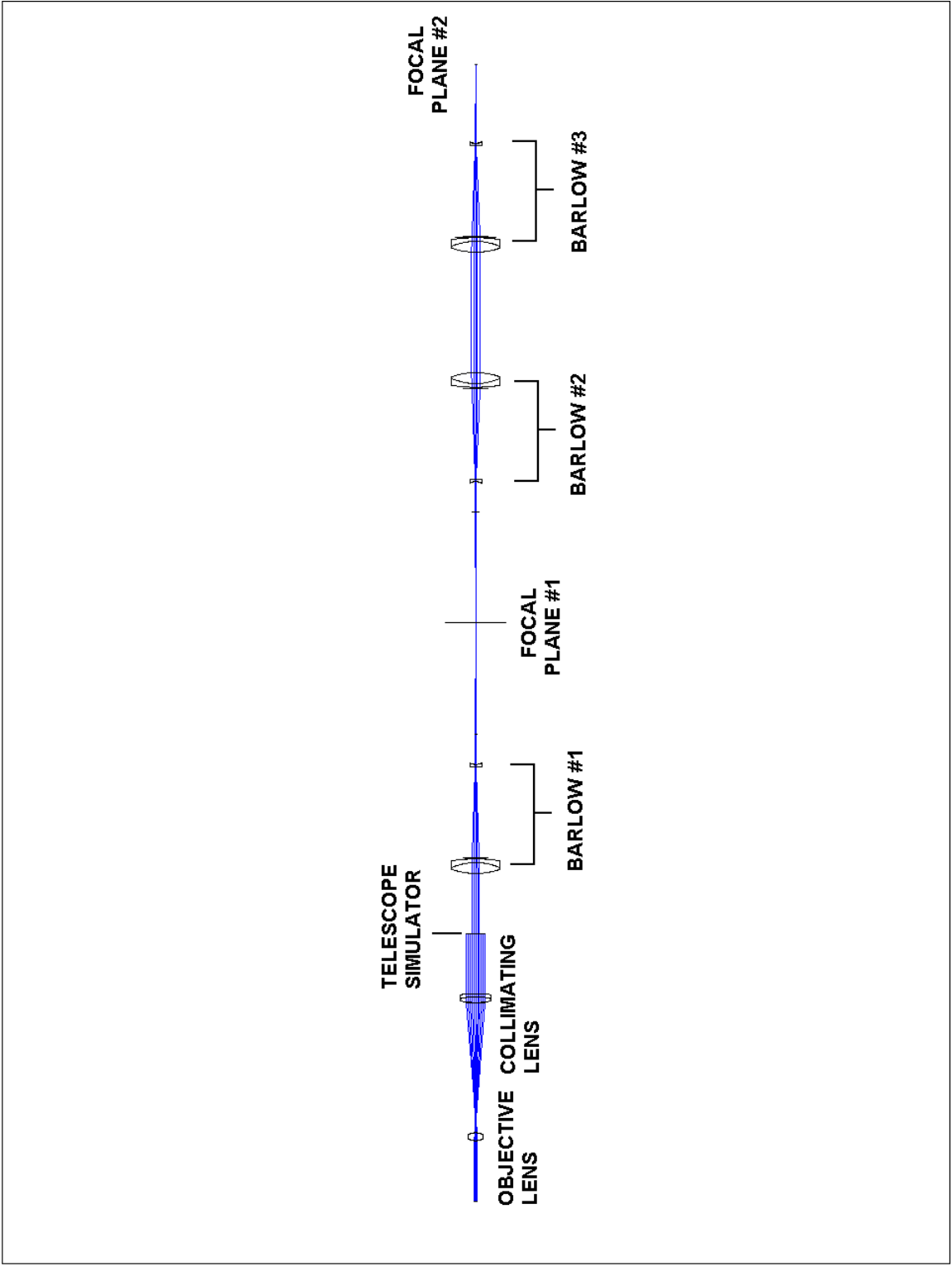

Fig.4(b)

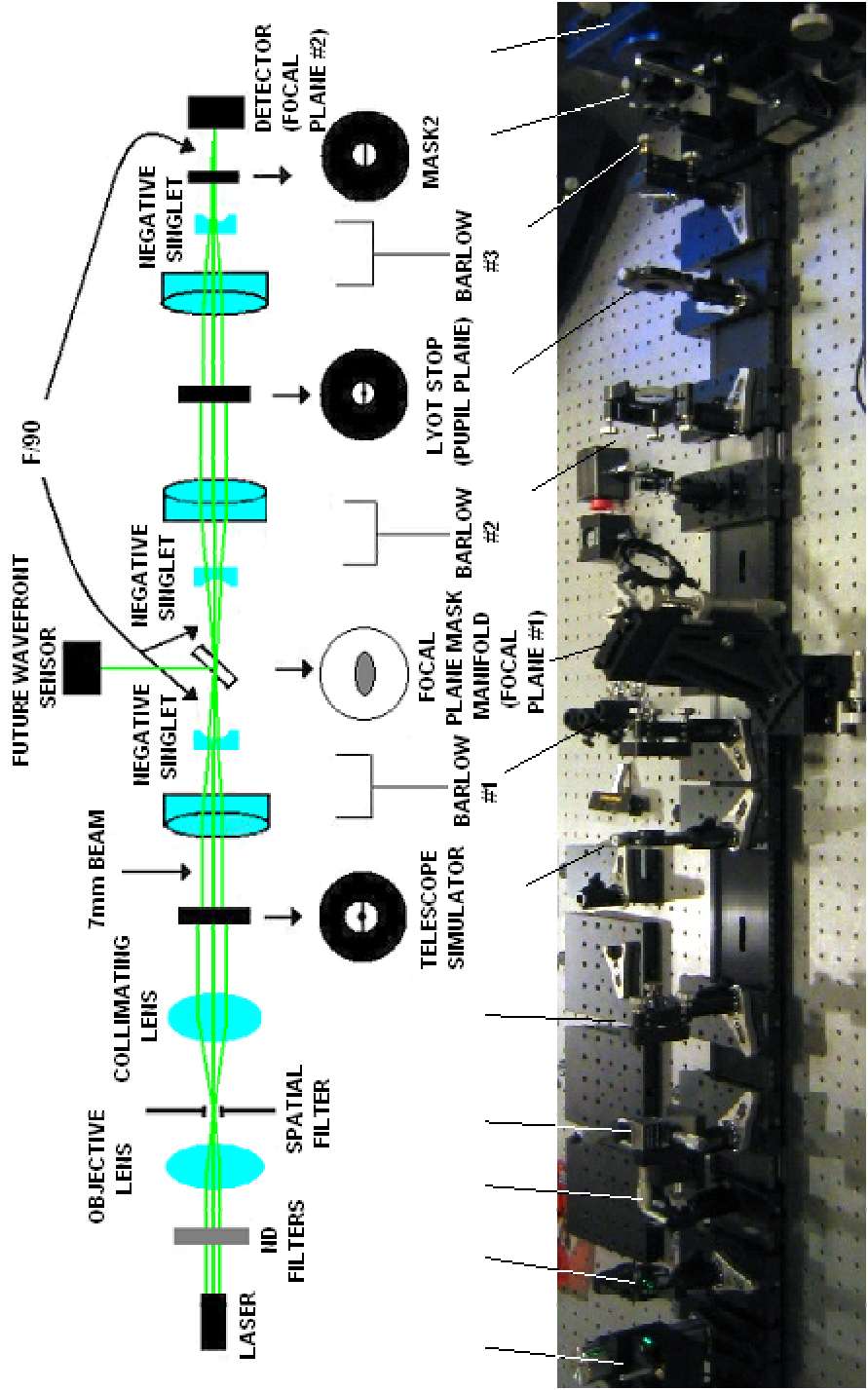

Fig. 5

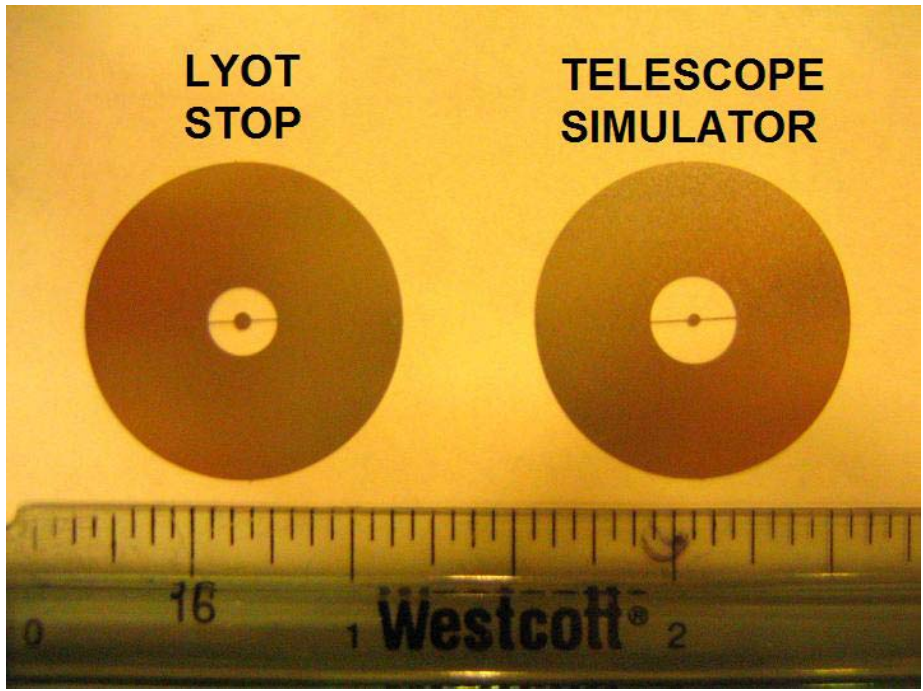

Fig. 6(a)

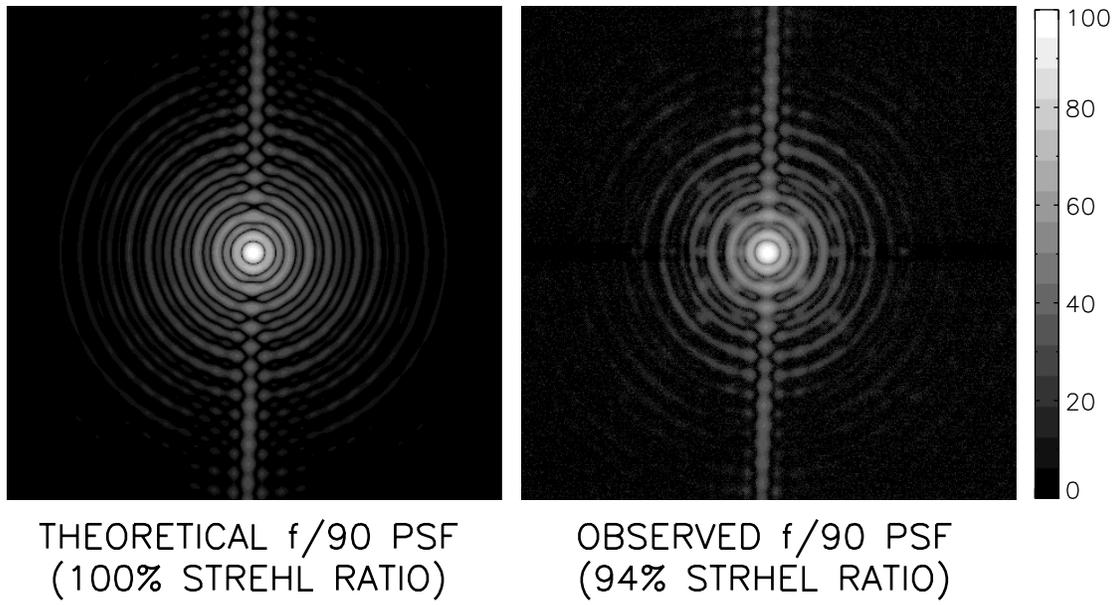

THEORETICAL f/90 PSF
(100% STREHL RATIO)

OBSERVED f/90 PSF
(94% STRHEL RATIO)

Fig. 6(b)

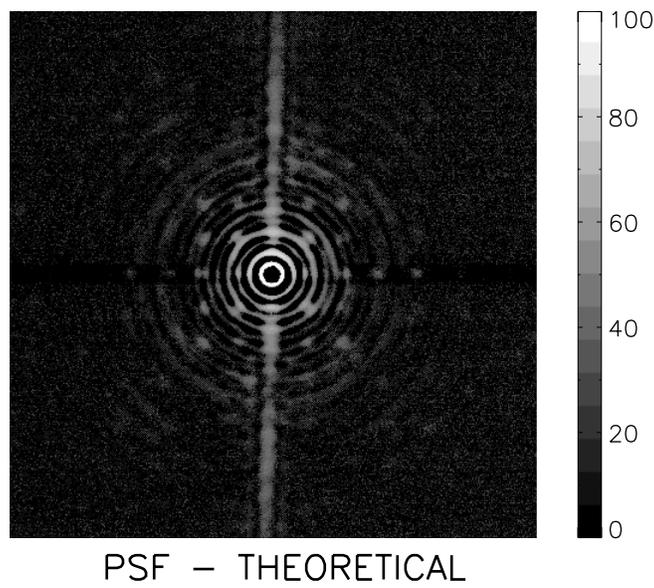

PSF – THEORETICAL

Fig. 6(c)

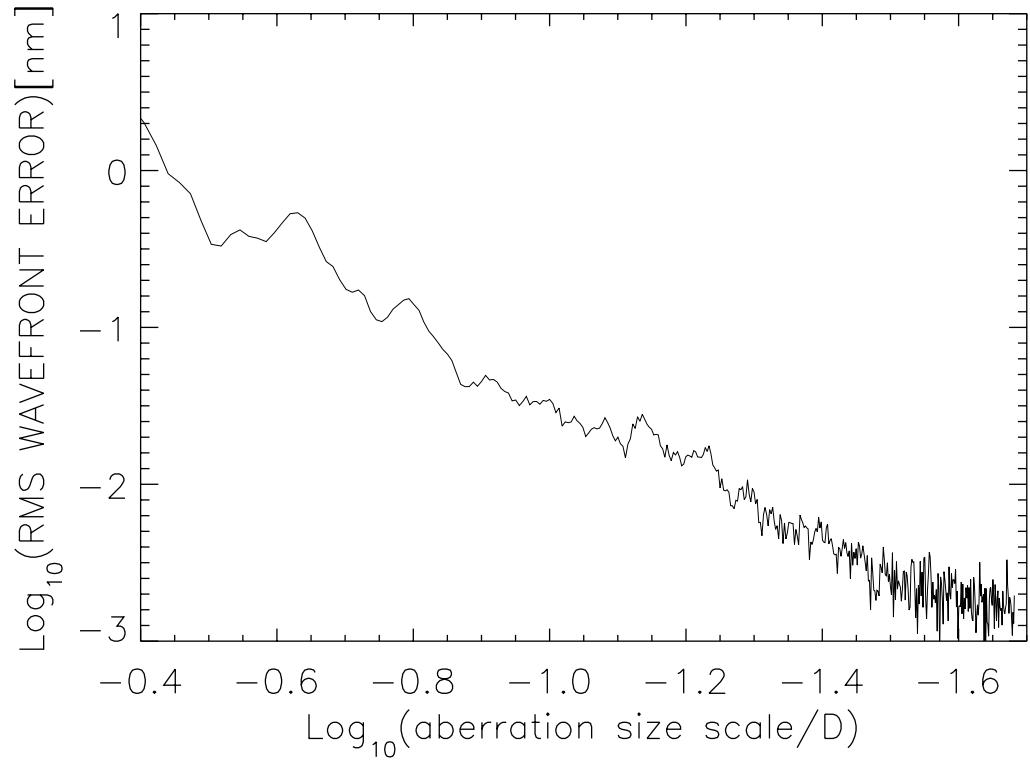

Fig.7(a)

(a) Mask size(r= 1.9 λ/D)

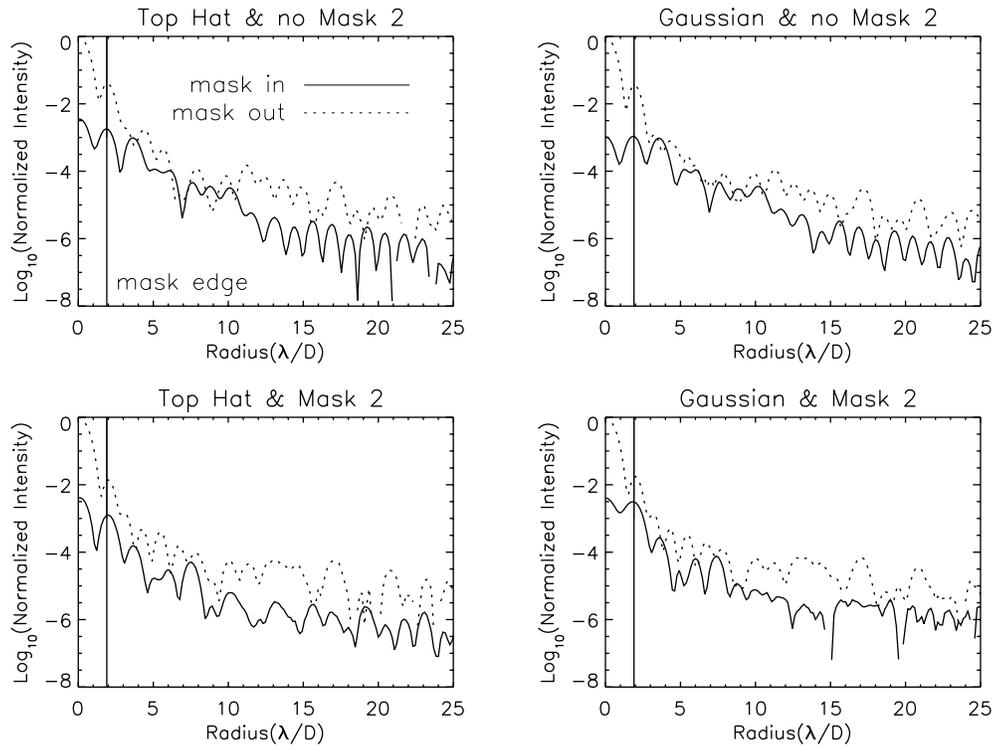

Fig. 7(b)

(b) Mask size(r= 3.7 λ/D)

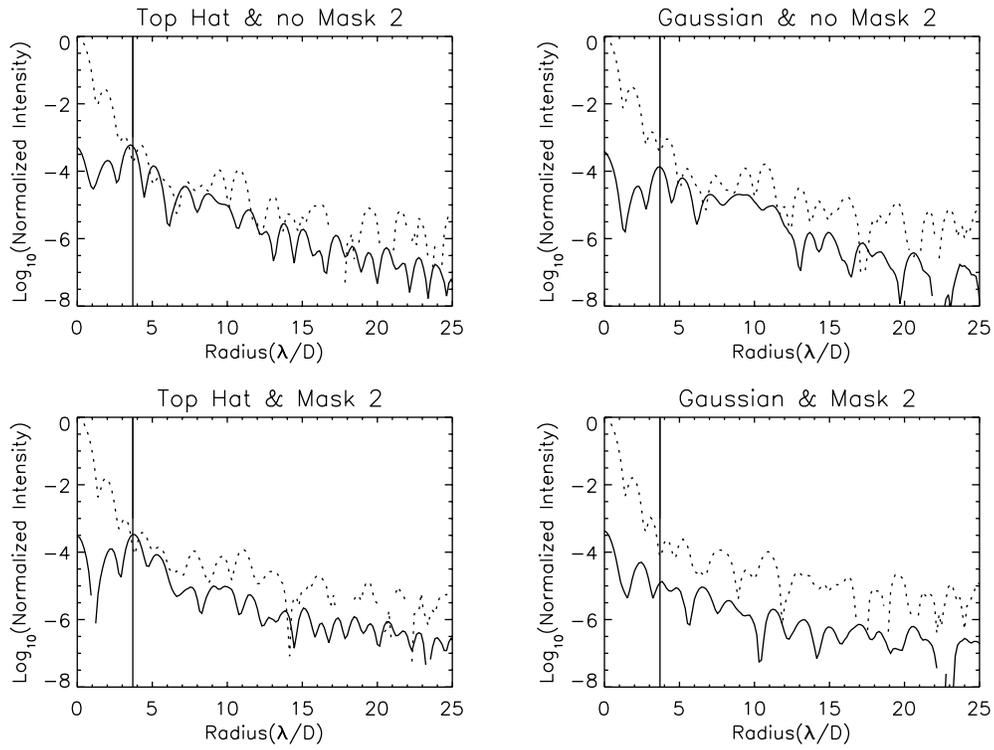

Fig. 7(c)

(c) Mask size(r= 7.4 λ/D)

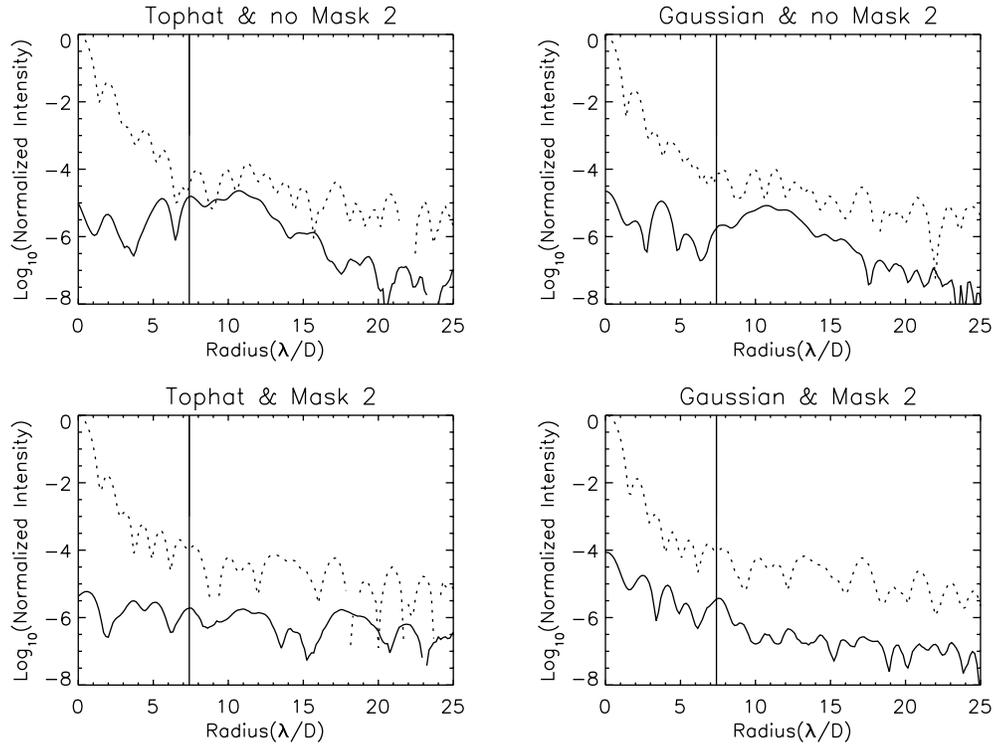

Fig. 8(a)

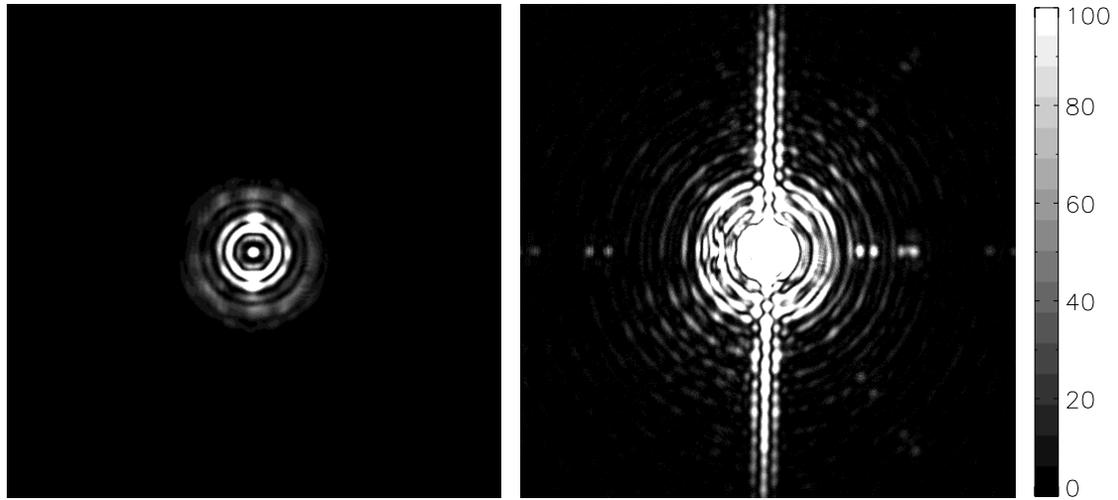

NO MASK 2
(FOCAL PLANE #1 MASK IN)                 NO MASK 2 (MASK OUT)

Fig. 8(b)

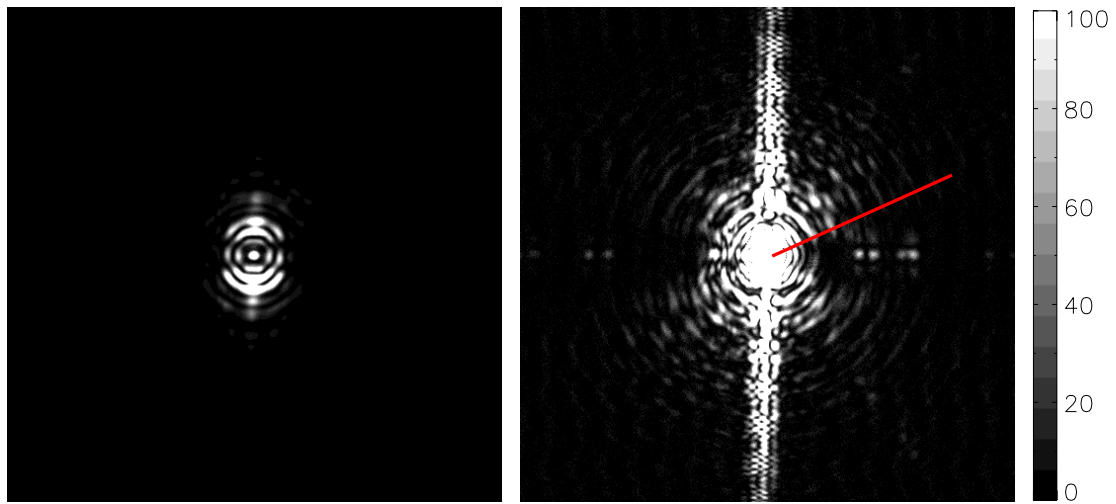

WITH MASK 2 (MASK IN)          WITH MASK 2 (MASK OUT)

Fig. 9

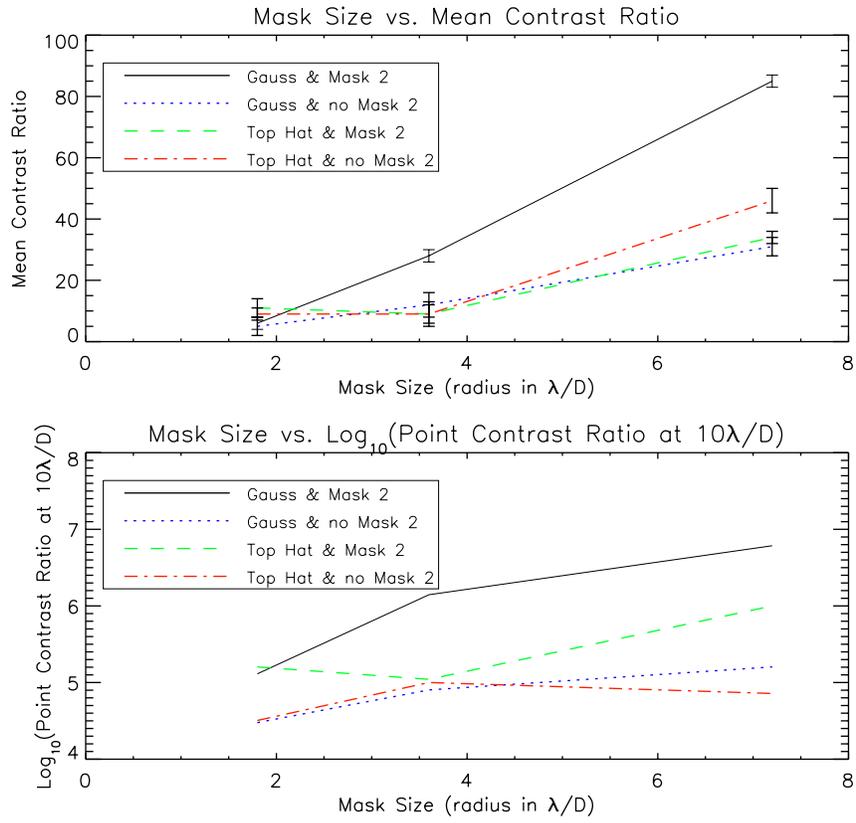

Fig. 10

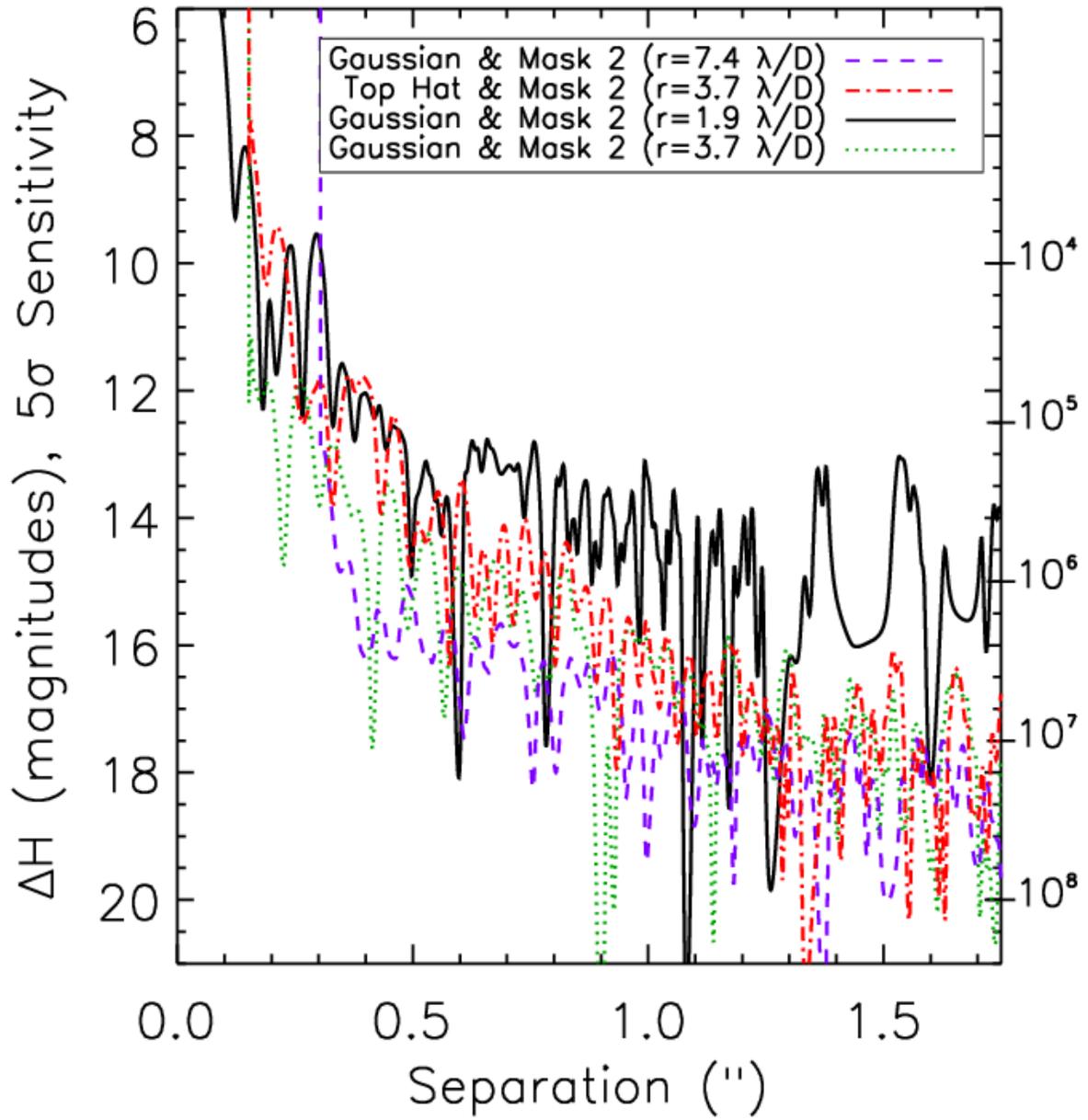

Fig.11(a)

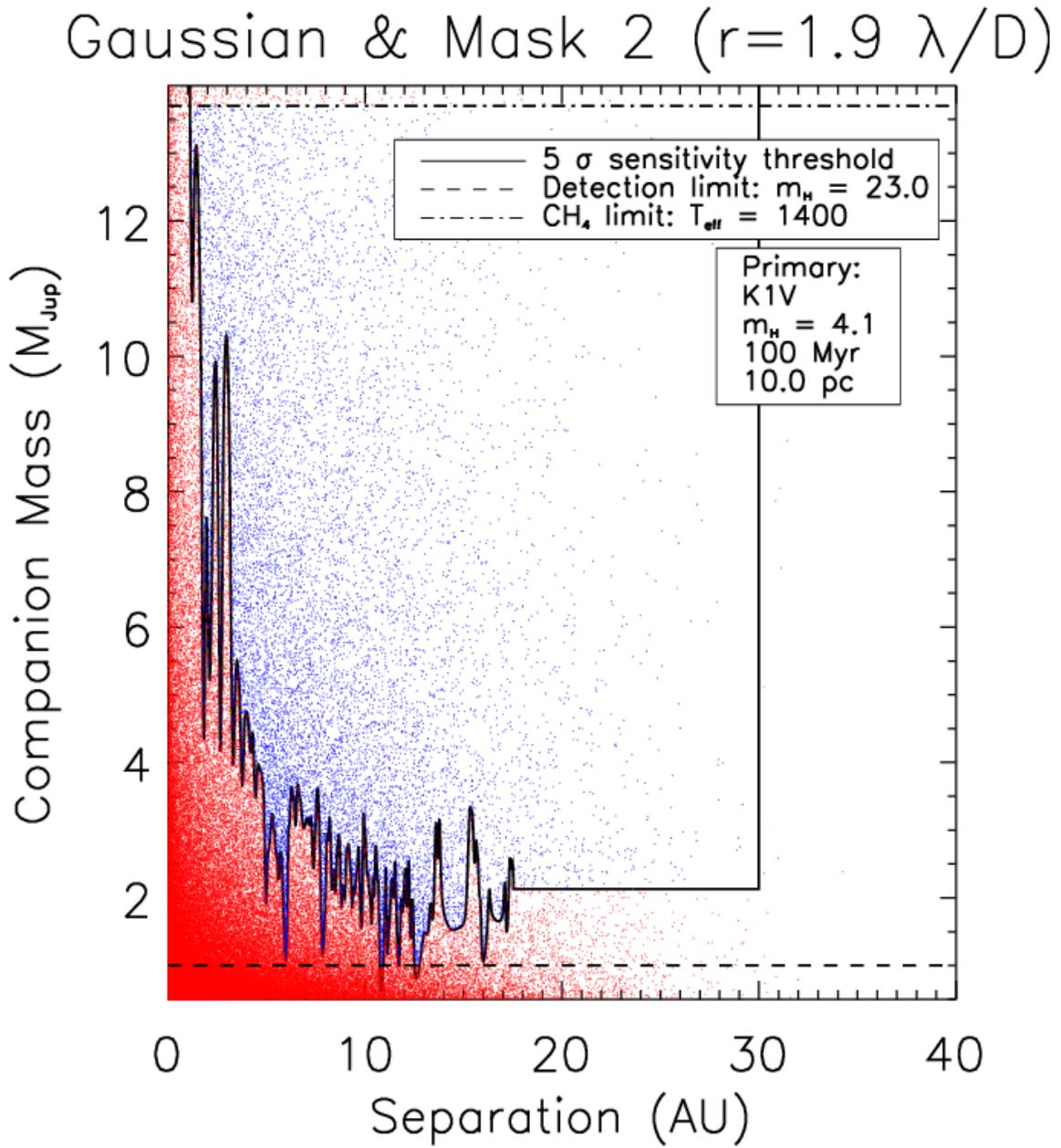

Fig. 11(b)

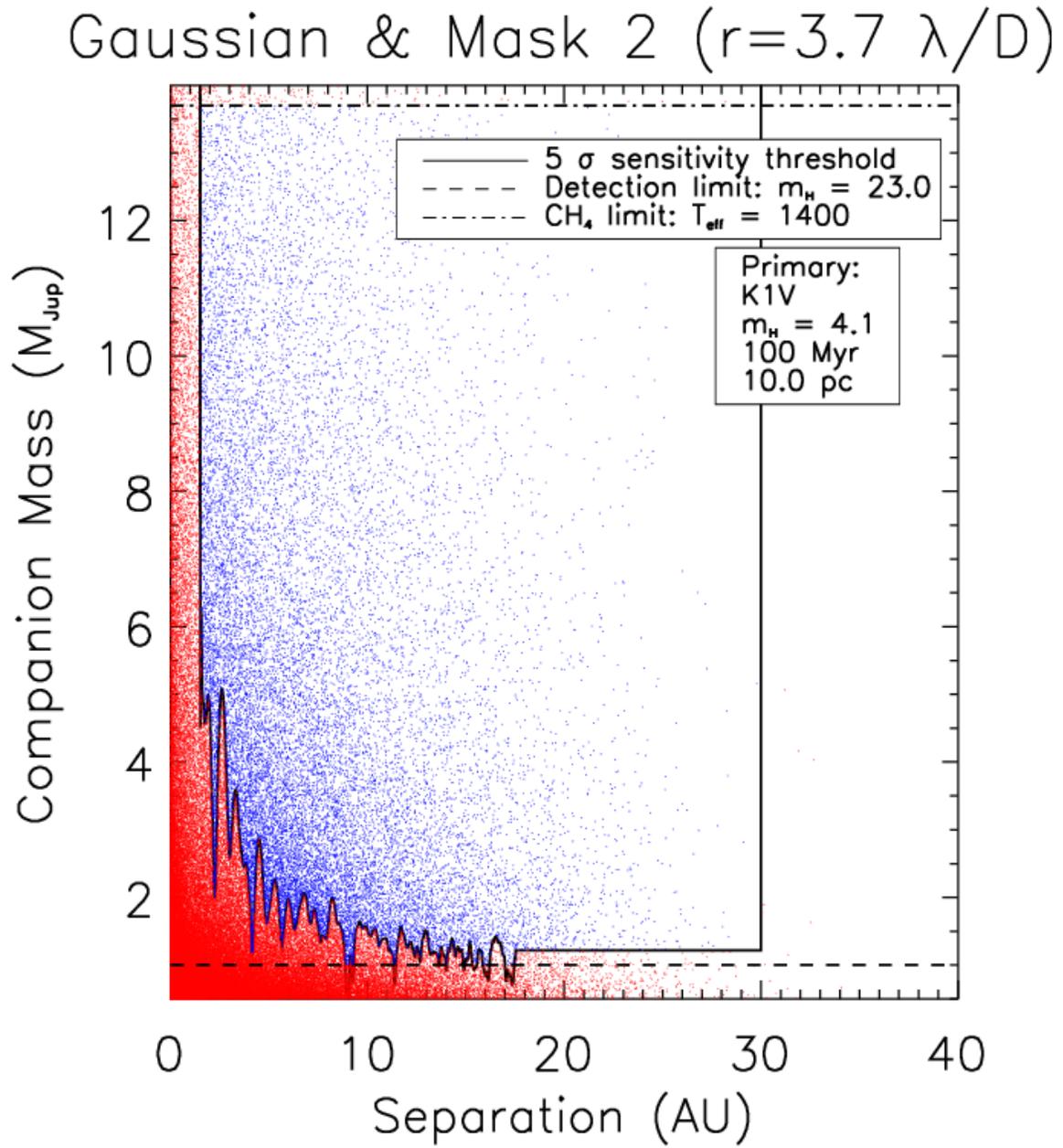

Fig. 11(c)

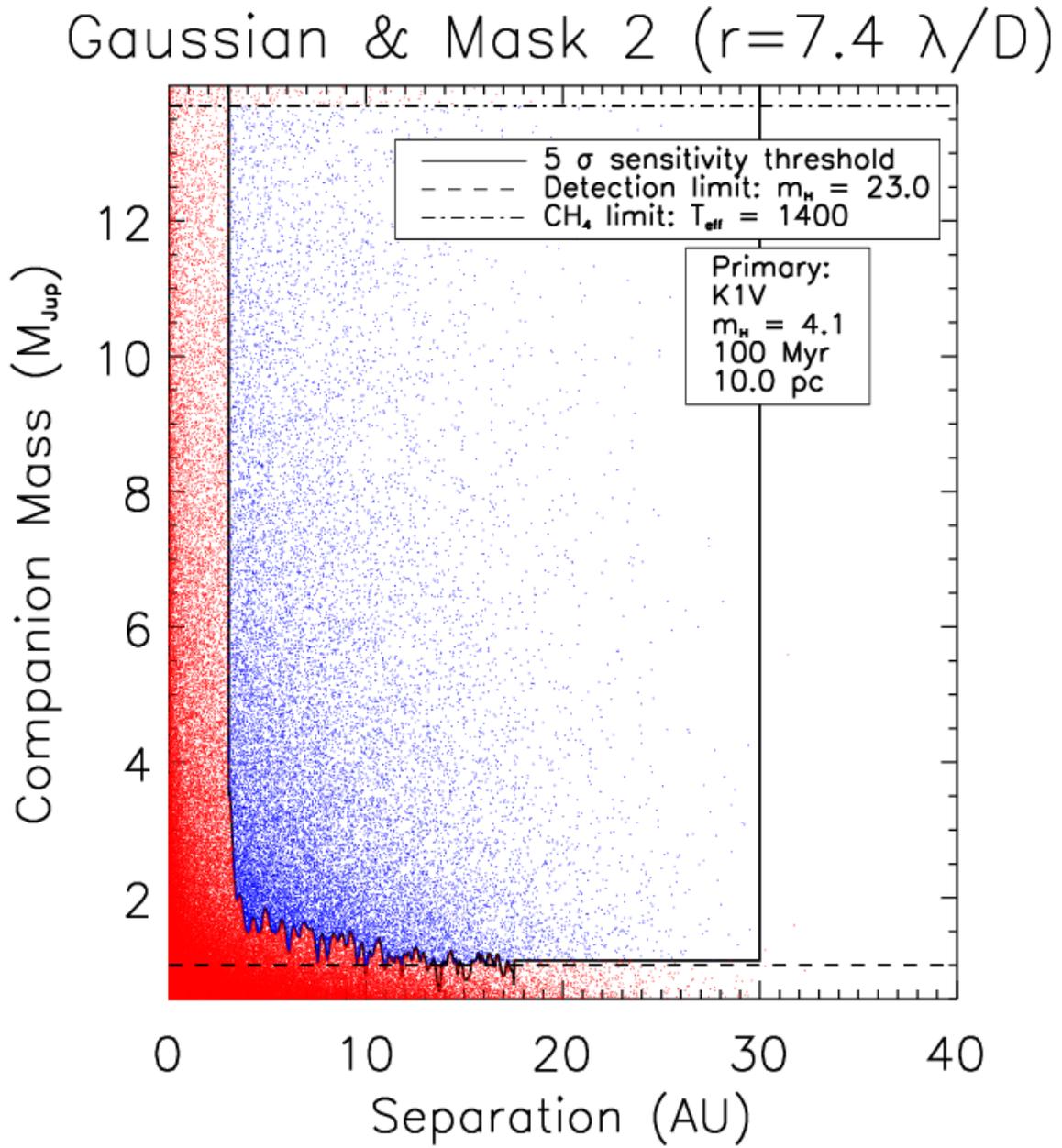

## Gaussian & Mask 2 (r=7.4 λ/D)

Fig. 11(d)

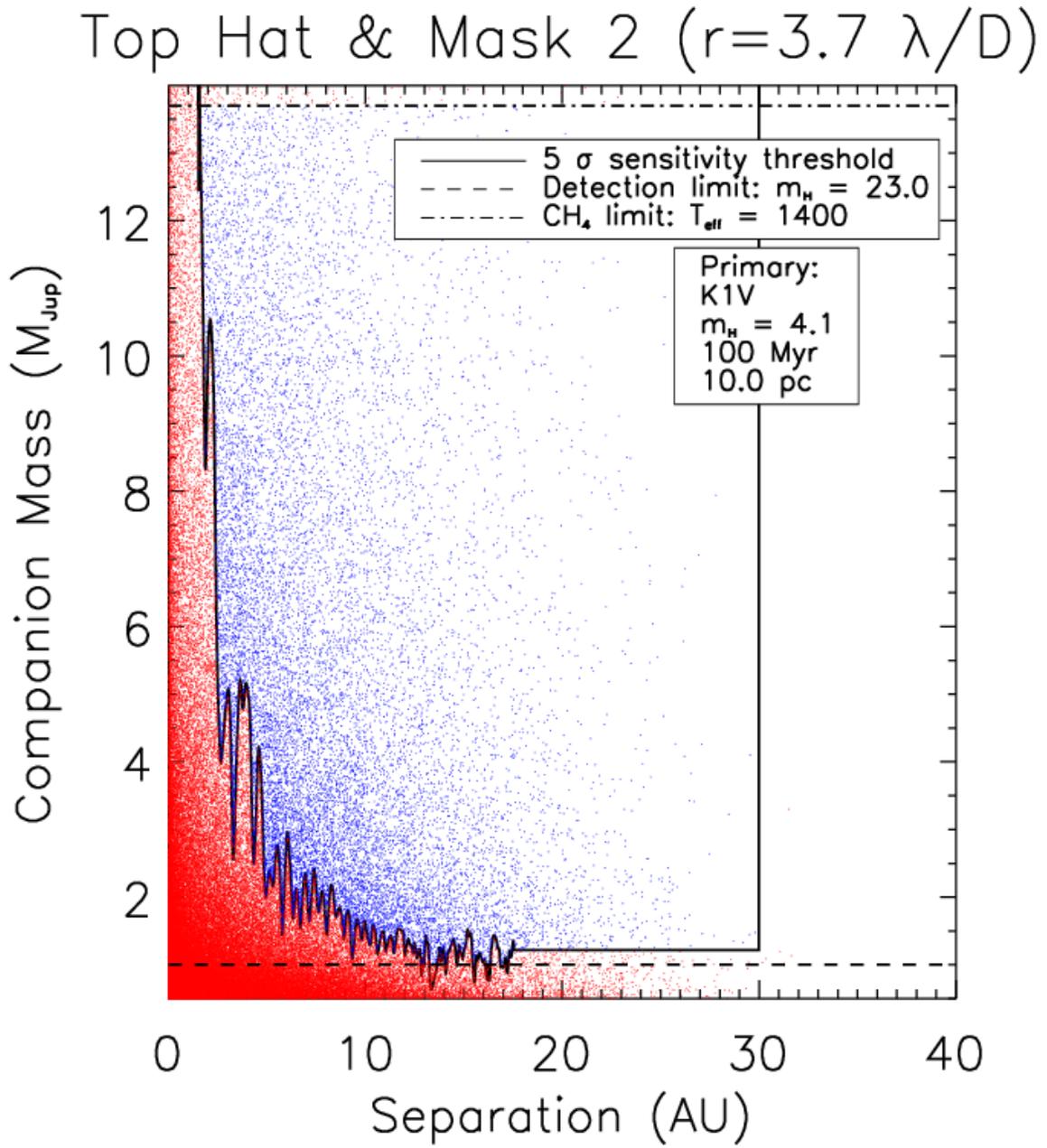

Top Hat & Mask 2 (r=3.7 λ/D)

Fig. 12.

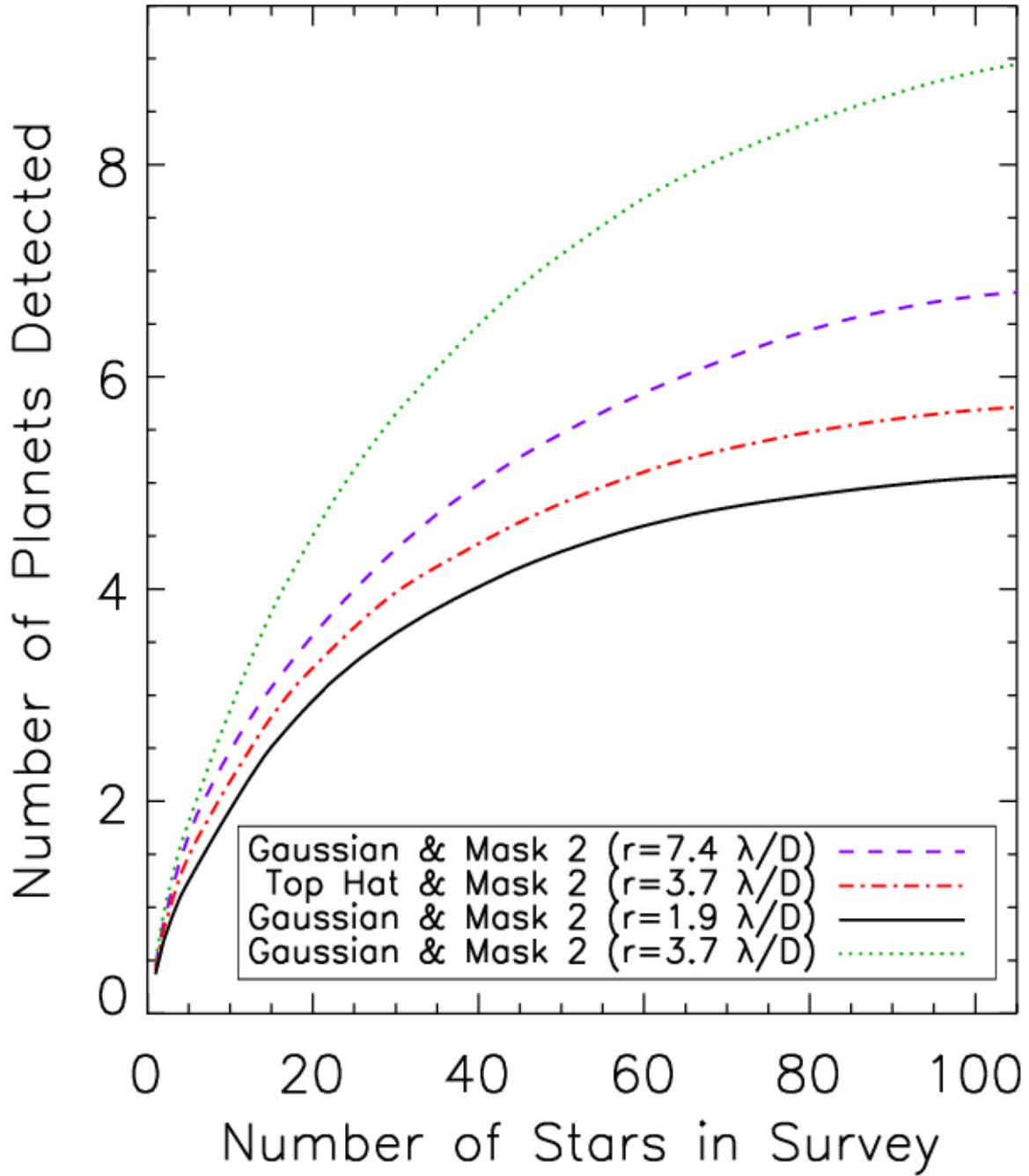